\documentclass[
    ,final            
  ]
  {aipproc}

\layoutstyle{6x9}

\begin{document}

\title{System Size Dependence of Azimuthal Correlations in Relativistic Heavy Ion Collisions 
}


\author{Wolf G. Holzmann, for the PHENIX Collaboration}{
  address={Department of Chemistry, SUNY Stony Brook, Stony Brook, NY 11794-3400, USA}
}

%

\begin{abstract}
Systematic comparisons of jet pair correlations obtained in Cu+Cu and Au+Au 
collisions at $\sqrt{s_{NN}}=$200~GeV are presented. The measured jet-pair 
distributions for both systems show strong modification of the away-side jet. 
For the same number of participating nucleons, the modification does not show 
a strong dependence on the collision system. It is suggested that such comparisons 
can provide important constraints for models which predict specific path length 
dependent jet modification effects.
\end{abstract}

\maketitle


\section{Introduction}
Accumulating experimental evidence suggests that matter with energy density in excess 
of what is required for the creation of a deconfined phase of quarks and gluons (QGP), 
is produced in Au+Au collisions at 
$\sqrt{s_{NN}}=$200~GeV at RHIC~\cite{Adcox:2004mh}.
Azimuthal correlation measurements provide an important probe for the properties 
of this matter since they allow the study of (di)jets. 
Hard scattered partons traversing hot and dense QCD matter can lose energy before 
fragmenting into hadrons. Such an energy loss results in significant modification
to jet properties in heavy ion collisions~\cite{bjorken,appel1986,wang2}. 
Indeed, the away-side jet in central and mid-central Au+Au collisions at $\sqrt{s_{NN}}=$200~GeV 
has been found to be strongly modified both in yield and shape~\cite{Adler:2002ct,star_fq,ppg032}.

	Jet modification is expected to be sensitive to several properties of the QCD medium, 
including the gluon density, formation time and the path length traversed by partons. 
Recently, much excitement has revolved
around discussions of jet-induced mach shocks~\cite{stoecker,colorwake}, 
primarily because of a possible sensitivity to the viscosity and the speed of sound in 
the nuclear collision medium. A crucial outstanding issue of great importance is the detailed
underlying mechanism for jet quenching. The resolution of this issue will undoubtedly 
rely on experimental constraints for the respective influence of energy density and path 
length on jet modification. Both vary with collision centrality.

	Two particle correlation measurements relative to the reaction plane, have been used in an initial
attempt to investigate path length effects~\cite{star_rp}. They indicate sensitivity  
to the orientation of the reaction plane~\cite{star_rp}. A complementary but simpler 
approach is to study jet-modification as a function of colliding system size.

It is well known~\cite{et} that the transverse energy per particle does not change significantly 
from SPS to RHIC energies ($\sqrt{s_{NN}}\approx~20-200$~GeV), suggesting that almost all of this
beam energy increase goes into particle production. Recent preliminary results from the PHOBOS 
collaboration also indicate that the measured charged particle pseudorapidity density in Cu+Cu and Au+Au 
are essentially identical [within errors] for the same number of participating 
nucleons ($N_{part}$). Taken together, these measurements suggest that a selection of the 
same $N_{part}$ in both collision systems targets matter with very similar energy density. 
On the other hand, the system sizes are different and should lead to different effective path 
lengths relevant to jet modification. Using a simple Glauber model and assuming an elliptic overlap
region, one can estimate system size differences of the order of 20-30\% for $N_{part} \approx 74$.
If triggering on high $p_T$ particles primarily selects jets from the surface of the fireball \cite{bias},
the differences in effective path lengths between both systems might be reduced.  
\section{Results and Summary}
The analysis presented here uses Au+Au and Cu+Cu data collected at the same 
collision energy ($\sqrt{s_{NN}}$=200 GeV) in the two central arms of 
the PHENIX detector. The detector setup is described elsewhere~\cite{nim_1}.
%
The data analysis is analogous to Ref.~\cite{ppg032}.

\begin{figure}[thb]
\vspace{-5.0cm}
\includegraphics[height=.3\textheight, width=17pc]{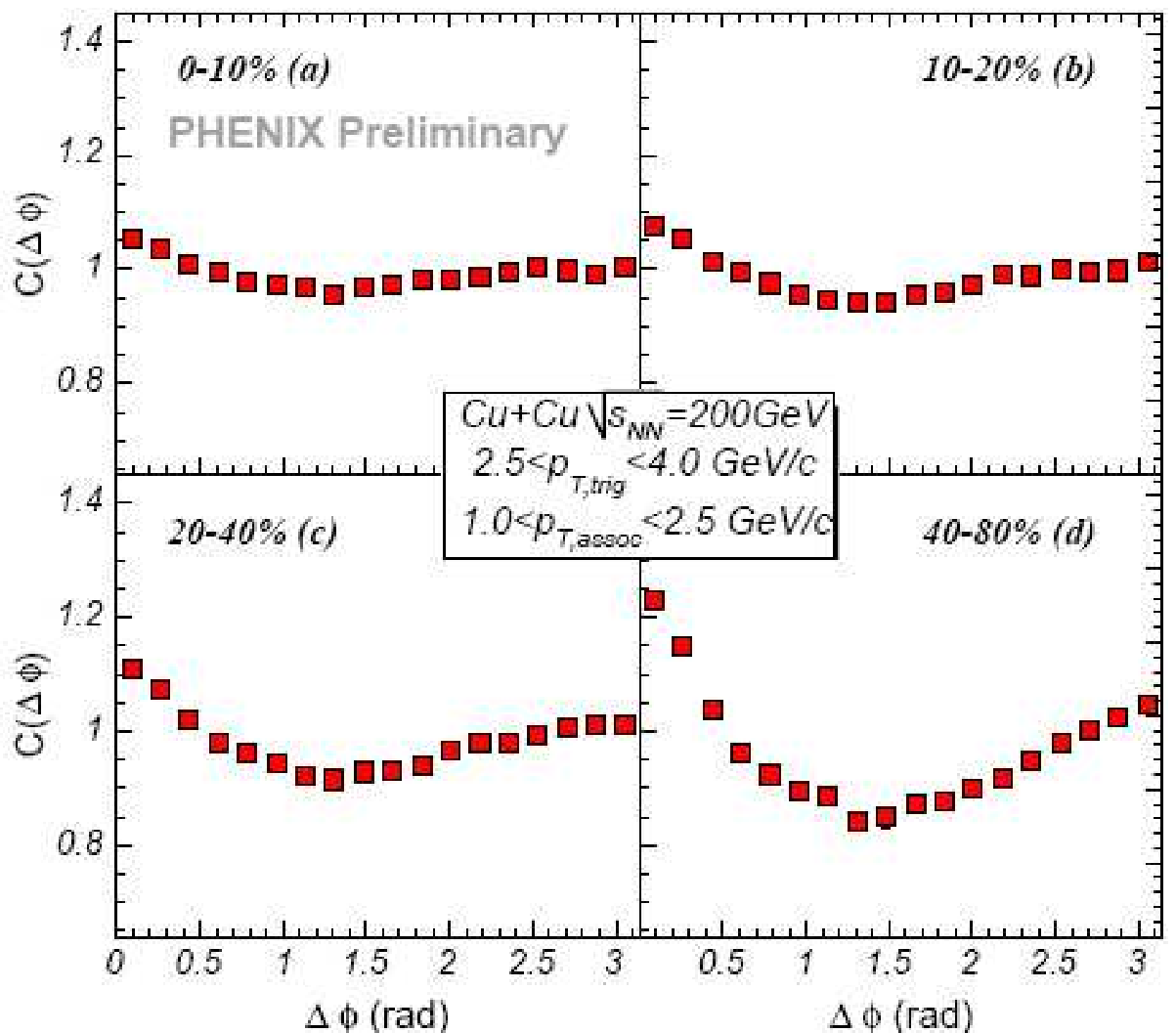}\vspace{-5.0cm}
\caption{Left: Correlation functions for several centrality selections in Cu+Cu as indicated. Right: Comparison of jet-pair distributions for Au+Au (blue points) and Cu+Cu (red squares) at $N_{part} \approx 74$.\label{fig:cf}}
\includegraphics[height=.25\textheight, width=17pc]{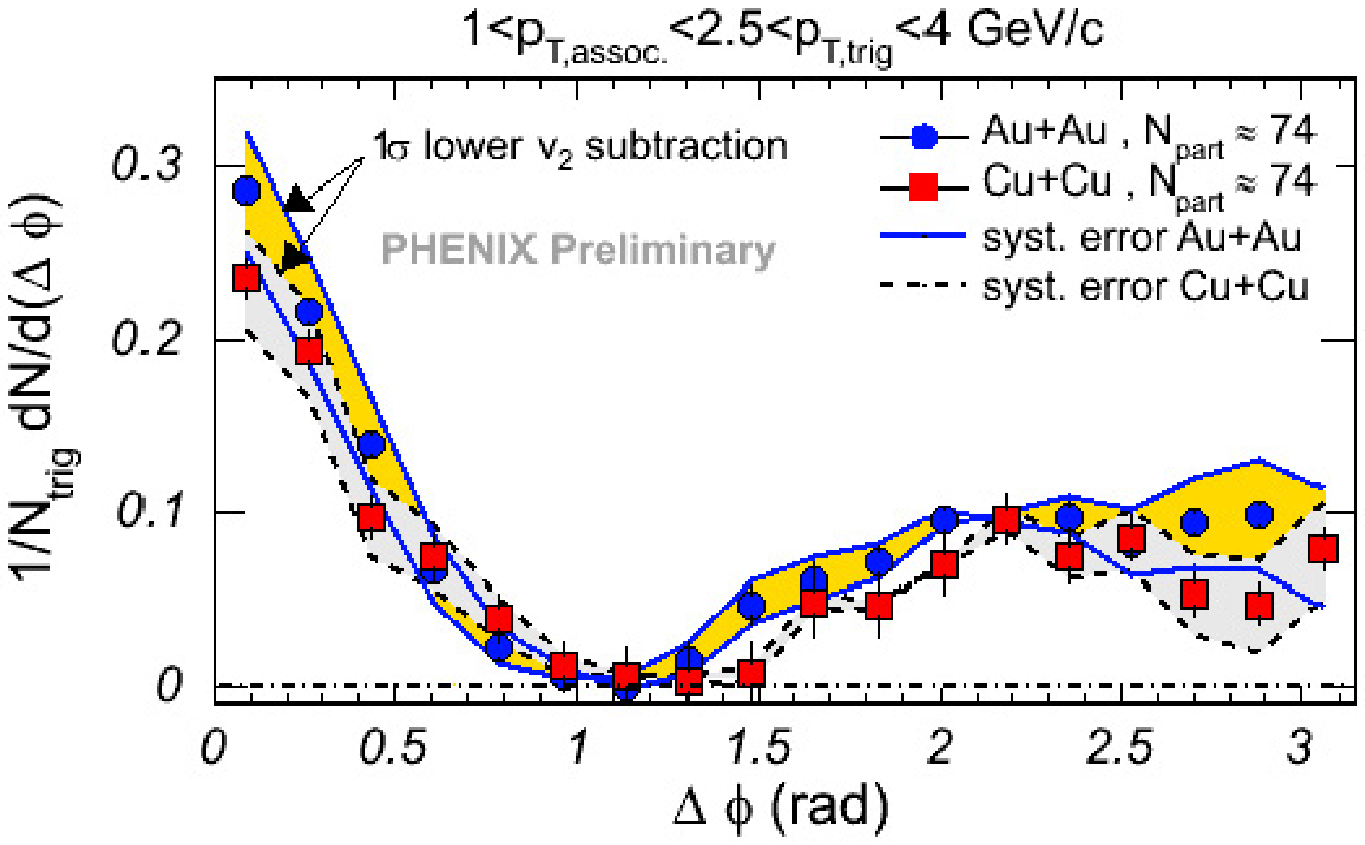}
\end{figure}

We define the correlation function as the ratio of two distributions in 
the azimuthal angle difference ($\Delta \phi$) between particle pairs formed with one hadron from a high-$p_{T}$ "trigger" bin ($2.5<p_T<4.0$ GeV/c) and another hadron from a lower "associated" $p_{T}$ selection ($1<p_T<2.5$ GeV/c).
A foreground distribution $N_{cor}$ is constructed with correlated particle pairs from the same event, and a background distribution $N_{mix}$ is obtained 
by randomly pairing particles from different events within the same multiplicity and vertex classes:
$C\left(\Delta\phi\right) \propto N_{cor}\left( \Delta\phi
\right) / N_{mix}\left( \Delta\phi \right)$. Fig.~\ref{fig:cf} (Left) shows correlation functions for several centrality selections in Cu+Cu collisions, as indicated.
%

For decomposition of the correlation functions, we assume that all correlation is only due 
to two sources: jet induced correlations and a harmonic component arising from elliptic flow. 
The harmonic amplitude is measured independently via a standard reaction plane analysis which 
minimizes non-flow correlations. This contribution is then subtracted from the correlation 
signal following the ZYAM constraint, detailed in Ref.~\cite{methods}. This procedure
gives the distribution of jet-correlated particle pairs per trigger particle.
%

\begin{figure}[thb]
  \includegraphics[height=.2\textheight]{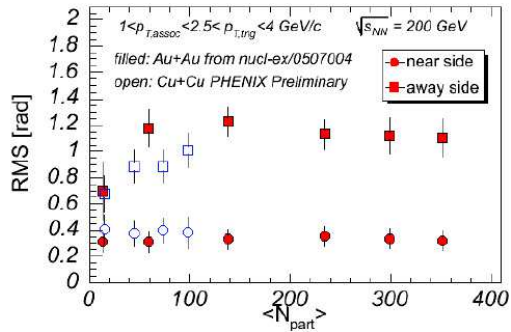} 
  \caption{Comparison of RMS widths from near- and away-side peaks for four centrality bins in Cu+Cu with    the same quantities from a recent Au+Au analysis \cite{ppg032}
  \label{fig:rms}}
\end{figure}

%
%
Jet pair distributions for Au+Au and Cu+Cu collisions (for $N_{part} \approx 74$)
are compared in Fig.~\ref{fig:cf} (Right). The two distributions show clear indications 
for significant broadening of the away-side jet. However, a strong system size dependence
is not evident, possibly suggesting that the energy density serves as the major actor in 
the jet modification process. The difference [albeit small] between the jet-pair distributions
for the Cu+Cu and the Au+Au systems serves to constrain models which predict strong path-length 
effects in the hot and dense collision medium. 
A complimentary view is provided in Fig.~\ref{fig:rms} which compares the RMS widths
extracted for the near- and away-side peaks for four separate centrality bins in Cu+Cu 
with a recent Au+Au analysis \cite{ppg032}. Although the away-side jet peaks are significantly broadened 
for mid-central and central Cu+Cu and Au+Au collisions, one can observe
a smooth $N_{part}$ dependence of the away-side widths for both systems. 
%

In summary, we have presented a systematic comparison of the charged hadron azimuthal
correlations between Au+Au and Cu+Cu collisions at $\sqrt{s_{NN}}=$200~GeV. The
jet pair distributions show significant broadening of the away-side jet in mid-central and 
central Cu+Cu and Au+Au collisions, but do not indicate a large system size dependence when  
the same $N_{part}$ and $p_T$ selections are made. This behavior
suggests that the energy density serves as the major actor 
for jet modification. Further studies to quantitatively pin down 
the detailed path-length dependence of jet-modification are currently underway.


\begin{thebibliography}{1}
%
\bibitem{Adcox:2004mh}
K. Adcox et al., \emph{Nucl. Phys.} \textbf{A757} (2005) 184


\bibitem{bjorken}J.D. Bjorken, (1982) FERMILAB-PUB-82-059-THY
\bibitem{appel1986}D.~A.~Appel, \emph{Phys. Rev.} \textbf{D33}, 717 (1986);
J.~P.~Blaizot et al., {\emph{Phys. Rev.} \textbf{D34}, 2739 (1986)
\bibitem{wang2} X.N. Wang, M. Gyulassy, \emph{Phys. Rev. Lett.} \textbf 68}, 1480 (1992); 
X.N.~Wang, \emph{Phys. Rev.} \textbf{C58}, 2321 (1998)


\bibitem{Adler:2002ct}C. Adler et al., \emph{Phys. Rev. Lett.} \textbf{90} 082302 (2003)
\bibitem{star_fq}J. Adams et al., \emph{Phys. Rev. Lett.} \textbf{95} 152301 (2005)
\bibitem{ppg032}S. S. Adler et al. [PHENIX Collaboration], nucl-ex/0507004

\bibitem{bias}A. Dainese, C. Loizides and G. Paic hep-ph/0511045

\bibitem{stoecker}H. St\"ocker, {\emph{Nucl. Phys.} \textbf{A750} (2005) 121;
J. Casalderrey-Solana et al., (2004) hep-ph/0411315
\bibitem{colorwake}J. Ruppert and B. M\"uller, \emph{Phys. Lett.} \textbf B618} (2005) 123-130 

\bibitem{star_rp}J. Adams et al., \emph{Phys. Rev. Lett.} \textbf{93} 252301 (2004)

\bibitem{et}S. S. Adler et al., \emph{Phys. Rev.} \textbf{C71}, 034908 (2005)

\bibitem{nim_1}K. Adcox et al., \emph{Nucl. Instrum. Meth.} \textbf{A499} (2003) 469
\bibitem{methods}N.N. Ajitanand et al., \emph{Phys. Rev.} \textbf{C72}, 011902 (2005)

%
%
%
%
\end{thebibliography}
\end{document}